\documentclass{IEEEcsmagArXiv}

\usepackage[colorlinks,urlcolor=blue,linkcolor=blue,citecolor=blue]{hyperref}
\expandafter\def\expandafter\UrlBreaks\expandafter{\UrlBreaks\do\/\do\*\do\-\do\~\do\'\do\"\do\-}
\usepackage{upmath,color}
\usepackage{todonotes}
\usepackage{hyperref}
\usepackage{tabularx, booktabs}
\usepackage{enumitem}

\jvol{XX}
\jnum{XX}
\paper{8}
\jmonth{~}
\jname{~}
\jtitle{Toward Research Software Categories}
\pubyear{2024}

\newcommand{\weg}[1]{#1}

\providecommand{\doi}[1]{\href{https://doi.org/#1}{#1}}

\newcommand{\ieeeitem}{\item[{\ieeeguilsinglright}]}

\begin{document}

\sptitle{~}

\title{Toward Research Software Categories}

\author{Wilhelm Hasselbring}
\affil{Software Engineering, Kiel University, Kiel, 24098, Germany}

\author{Stephan Druskat}
\affil{German Aerospace Center (DLR), Berlin, 12489, Germany}

\author{Jan Bernoth}
\affil{University of Potsdam, Potsdam, 14476, Germany}

\author{Philine Betker}
\affil{Department for Epidemiology, Helmholtz Centre for Infection Research, Brunswick, Germany} 

\author{Michael Felderer}
\affil{German Aerospace Center (DLR) \& University of Cologne, Cologne, 51147 , Germany}

\author{Stephan Ferenz}
\affil{Department of Computer Science, Carl von Ossietzky Universität Oldenburg, 26129 Oldenburg}

\author{Anna-Lena Lamprecht}
\affil{University of Potsdam, Potsdam, 14476, Germany}

\author{Jan Linxweiler}
\affil{TU Braunschweig, Braunschweig, 38106, Germany}

\author{Bernhard Rumpe}
\affil{Software Engineering, RWTH Aachen University, Germany}

\begin{abstract}
Research software has been categorized in different contexts to serve different goals.
We start with a look at what research software is, before we discuss the purpose of research software categories. We propose a multi-dimensional categorization of research software. We present a template for characterizing such categories. As selected dimensions, we present our proposed role-based, developer-based, and maturity-based categories. Since our work has been inspired by various previous efforts to categorize research software, we discuss them as related works. We characterize all these categories via the previously introduced template, to enable a systematic comparison.
\end{abstract}

\maketitle

\chapteri Research software is software that is designed and developed to support research activities. 
Research software is developed by researchers themselves or by software engineers working closely with researchers.
Research software is typically developed to meet specific research needs, and often has unique requirements that are different from standard commercial software~\cite{CiSE2018}.
However, research software is gaining appreciation and endorsement for research and as a research result itself
\weg{\cite{JayEtAl2021,AnztEtAl2021}}.

Research Software Engineering (RSE) is a specialized field that applies software engineering principles to address the unique challenges posed by developing software for scientific and academic research, with the goal of enhancing the efficiency, reproducibility, and impact of research outcomes. Research software engineers specialize in developing and maintaining software for scientific research purposes. 

In this paper, we propose a multi-dimensional categorization of research software, along the dimensions of roles, developers, and maturity.
We start with a look at what research software is before we discuss the purpose of research software categories. We present a template for characterizing such categories. Subsequently, our proposed role-based, developer-based, and maturity-based categories are presented.
Our work has been inspired by various previous efforts to categorize research software, which we discuss as related works. We characterize all these categories via the previously introduced template, and conclude with an outlook to future work.

\section{Research Software}

For the purposes of this paper, we follow the FAIR for Research Software (FAIR4RS) Working Group in their definition of research software, as software that was created during the research process or for a research purpose~\cite{FAIR4RS2022}
\weg{\cite{GruenpeterEtAl2021}}.
This \textit{prescriptive} definition distinguishes ``research software'' and ``software in research,'' which includes general purpose software.
The software components (e.g., operating systems, programming languages, libraries, etc.) that are used for research but were \emph{not} created during or with a clear research intent should be considered ``software in research'' and not ``research software.''

\begin{figure}[hb]
	\includegraphics[width=\linewidth]{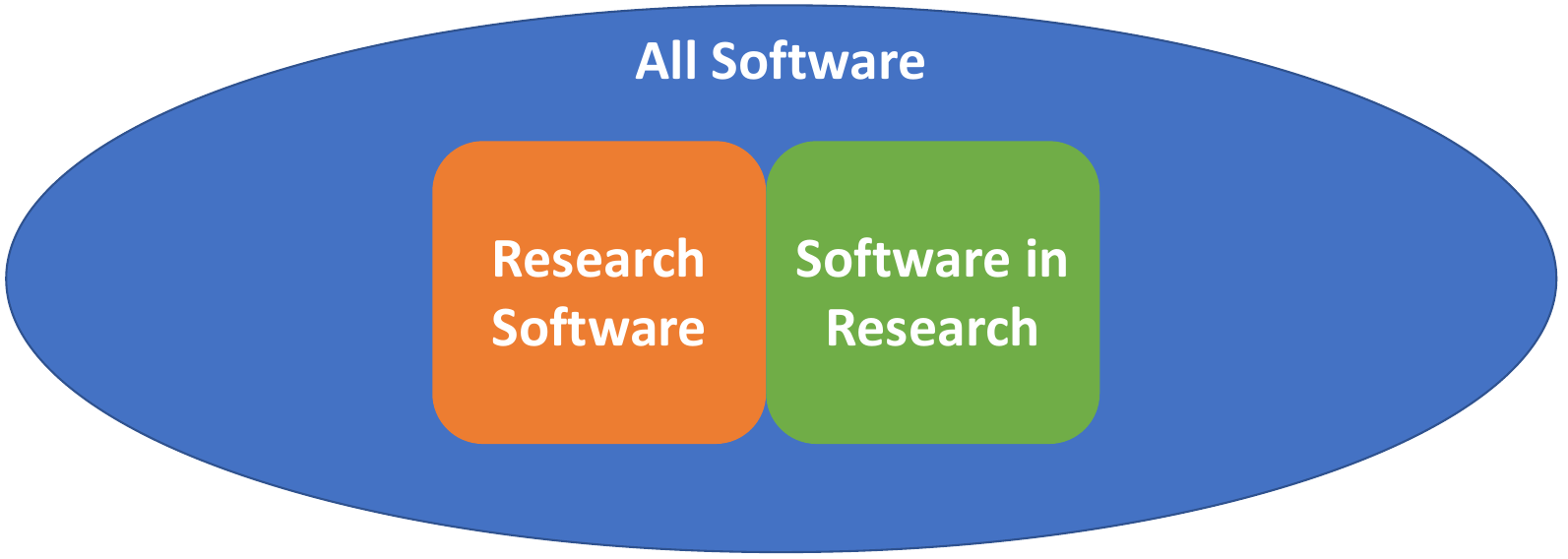}
	\caption{Segmentation of all software, research software, and software in research. In the present paper, we further categorize the orange box, i.e., research software.}
	\label{fig:SoftwareSegmentation}
\end{figure}

A \textit{descriptive} definition of research software could instead include all the software used in research, as for instance done in~\cite{OSRS2020} for analyzing GitHub repositories.
While such a descriptive definition may be useful in analyzing research processes, and therefore may be useful for RSE research~\cite{RSER2023}, the prescriptive definition defines a clearer focus for the work presented here, and enables a better disambiguation of properties specific to research software. 
The Research Data Alliance also adopted the prescriptive distinction between \emph{research software} and \emph{software in research}~\cite{FAIR4RS2022}, as we do. \autoref{fig:SoftwareSegmentation} shows the resulting segmentation of software.

\section{Purpose of Research Software Categories}

We envision the following benefits from using categories for research software, which may serve
    \begin{itemize}
        \ieeeitem as a basis of institutional guidelines and checklists for research software development;
        \ieeeitem to better understand the different types of research software and their specific quality requirements;
        \ieeeitem to recommend appropriate software engineering methods for the individual categories;
        \ieeeitem to design appropriate teaching / education programs for the individual categories;
        \ieeeitem to give stakeholders (especially research software engineers and their management) a better understanding of what kind of software they develop;
        \ieeeitem for a better assessment of existing software when deciding to reuse it;
        \ieeeitem for research funding agencies, to define appropriate funding schemes;
        \ieeeitem to define appropriate metadata labels for FAIR research software\weg{~\cite{Lamprecht2020,FAIR_Software_2020}};
        \ieeeitem in RSE Research~\cite{RSER2023}, to provide a framework for classifying research software artifacts.
    \end{itemize}
This list is not exhaustive.

\section{Characterization of Research Software Categories}

Categorizations can be described through their scope, purpose, context, properties, consequences for creation and use, and their inter-categorial relations. \autoref{table:categorization-criteria} provides a template for systematically describing the characteristics of research software categorizations, which we will use later to characterize some individual categories in the subsequent sections.

\begin{table*}[bht]
    \centering
    \begin{tabularx}{\textwidth}{l X}
    \toprule
        Criterion & Explanation \\
    \midrule
        Scope & What is the scope of the categorization? \\
        Purpose & What is the purpose of the categorization? \\
        Context &  In which contexts are specific categories developed and used? \\
        Properties & What are specific properties of the different categories? \\
        Consequences for Creation & How is and should software of a specific category be developed? \\
        Consequences for Use & How and why is software of a specific category used? What are the differences between the categories in terms of use and reuse, including, e.g., in software publication \& citation? \\
        Inter-categorial relations & What are the relations between different categories?\\
    \bottomrule
    \end{tabularx}
    \caption{Template for describing criteria of research software categorizations.}
    \label{table:categorization-criteria}
\end{table*}

\section{Role-Based Categorization of Research Software}

Research software can be used to collect, process, analyze, and visualize data, as well as to model complex phenomena and run sophisticated simulations. Research software is also developed to control and monitor lab experiments and environmental observations. In engineering research, research software meanwhile constitutes a new paradigm of scientific inquiry next to theory and experiment\weg{~\cite{hey2009the}} and acts as a proof-of-concept to invent and evaluate new technological artifacts, including algorithms, methods, systems, tools, and other computer-based technologies. Research software also provides the infrastructure to manage, publish, and archive research data and software. 

Thus, research software may take various \emph{roles} in the research process~\cite{NieuwpoortKatz2023}.
This is similar to software engineering teams, which involve a range of roles that contribute to the development, maintenance, and improvement of software systems. Some common roles in software engineering are software architect, programmer, and tester. Each role may be taken by several persons, and one person may take several roles. These role assignments may also change during a software project. 

We propose a similar role-based categorization of research software, with an emphasis on varying quality requirements for the different \emph{roles}, which software may take in research.
Accordingly, a research software may take several roles, which may also change during the life cycle of the software.

Research software mainly falls into one of the following three top-level role categories (and sometimes combinations):
\begin{enumerate}
	\item \emph{Modeling, Simulation, and Data Analytics} of, e.g., physical, chemical, social, or biological processes in spatio-temporal contexts.
	\item \emph{Proof-of-Concept Software} in science and engineering research.   
	\item \emph{Research Infrastructure Software}, such as research data and software management systems.
\end{enumerate}
The assignment of a research software to categories may evolve over time. For instance, software specifically developed for a research question (usually Categories 1 \&\ 2) can later turn into infrastructure software (Category~3)\weg{~\cite{Katz2022}}.
In different contexts, a software may also be in multiple categories at the same time.
As an example, the Monticore~\cite{Krahn2010} framework for the development of domain specific languages initially was a proof-of-concept for researching implementation methods for developing domain specific languages. Meanwhile, Monticore is mainly used for developing specific domain specific languages, thus it turned into a research infrastructure. Still, some original research on developing domain specific languages is done in the Monticore context.

Category~1) software for modeling, simulation, and data analytics is quite large. We further refine this category with several subcategories:

\begin{enumerate}[label=1.\arabic*)]
\itemindent=1em
	\item Modeling and simulation (e.g., numerical modeling, agent-based modeling).
	\item Data analytics, on observation and simulation data, with statistical analysis and machine learning as methods.
	\item Integrative analysis (data assimilation and decision analysis)
	\item Scientific visualization
\end{enumerate}
Category~2) for proof-of-concept software is used in structural sciences (mathematics and computer science) and in engineering sciences (software, electrical, mechanical, and civil engineering). 
\weg{Examples for proof-of-concept research software are a software for flexible energy management of vehicle fleets within a harbor terminal that was developed and evaluated in the FRESH research project~\cite{holly_flexibility_2020}, and the Kieker monitoring framework that has been employed in various software engineering research projects~\cite{Kieker2020}.}

Category~3) for research infrastructure software is also quite large. We further refine this category with several subcategories:
\begin{enumerate}[label=3.\arabic*)]
\itemindent=1em
	\item Control and monitoring software for complex experiments and instruments. This includes embedded control software, as well as native and web-based monitoring software.
	\item Data collection (survey software, sensor-based data collection, etc.).
	\item Pipelines, Workflows, and frameworks for composing software components.
	\item Libraries, for instance for high performance computing.
	\item Laboratory Notebooks.
	\item Data Management.
	\item Software Management.
	\item Collaboration software.
\end{enumerate}
These categories have varying quality requirements. 
For instance, dedicated requirements engineering may be relevant for Category~3), but not for Category~1).
As another example, safety analysis may be relevant for Category~3.1), but not for Categories~1) and~2).

\begin{figure*}[p]
	\includegraphics[width=\textwidth]{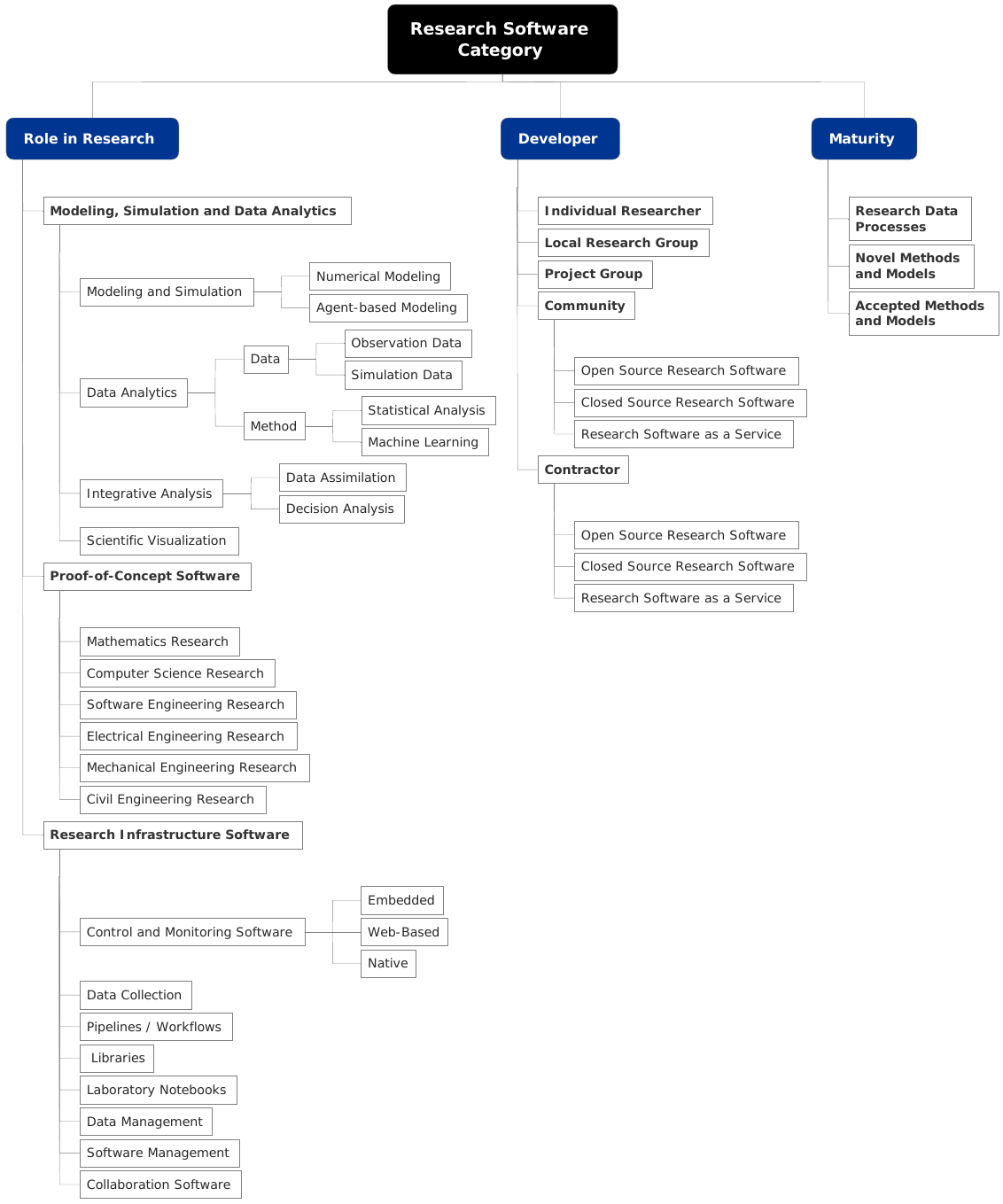}
	\caption{Our multi-dimensional categorization of research software, along the dimensions of roles, developers, and maturity.}
	\label{fig:SoftwareCategorization}
\end{figure*}

\autoref{fig:SoftwareCategorization}, left, shows our resulting role-based categorization.

\autoref{table:Our-categorization} characterizes our multi-dimensional categorization in terms of the template in \autoref{table:categorization-criteria}.
The developer-based and maturity-based categorizations are introduced in the following two sections, before we discuss some related categorizations.

\begin{table*}[bt]
    \centering
    \begin{tabularx}{\textwidth}{l X}
    \toprule
        Criterion & Explanation \\
    \midrule
        Scope & This categorization covers the dimensions of roles, developers, and maturity. \\
        Purpose & The categorization aims to enable a better understanding of the different types of research software and their specific quality requirements.\\
        Context & The categorization has been produced in the context of a task force of the special interest group on Research Software Engineering, within the German Association of Computer Science (GI e.V.) and the German Society for Research Software (de-RSE e.V.). It is meant to serve different purposes, in particular RSE research~\cite{RSER2023}. \\
        Properties & The categories follow different relevant dimensions, and are defined collaboratively among software engineering researchers and research software engineers. \\
        Consequences for Creation & Depending on its category, software is expected to meet different quality requirements and follow different development processes.\\
        Consequences for Use &  Perceive that there are many different types of research software, fulfilling many different roles and functions.\\
        Inter-categorial relations & Individual research software may change its category within one or more dimensions.\\
    \bottomrule
    \end{tabularx}
    \caption{Characteristics of our multi-dimensional categorization for research software.}
    \label{table:Our-categorization}
\end{table*}

\section{Developer-Based Categorization of Research Software}

For the developer dimension, we see the following stages for research software:
\begin{enumerate}
	\item Individual Researcher, such as PhD student or PostDoc.
	\item Local Research Group
	\item Project Group, in which several research groups may collaborate.
	\item Community on a specific research topic.
	\item Contractor (professional software company developing the software on behalf of the researchers).
\end{enumerate}
A community or contractor may develop the software open-source, closed-source, or it may provide research software as an online service. An example for a contractor-developed research software is PIA, the Prospective Monitoring and Management App, which has been developed by professional software companies as open-source software, on behalf of the Helmholtz Institute for Infection Research to conduct observational epidemiological studies by facilitating longitudinal data collection and cohort management ~\cite{HEISE2022100931}.

\autoref{fig:SoftwareCategorization}, middle, shows our resulting developer-based categorization.

\section{Maturity-Based Categorization of Research Software}

Concerning a maturity-based categorization of research software, we adopt the ARDC approach~\cite{ARDC2022}:
\begin{enumerate}
	\item \emph{Research Data Processes} captured as software. The result is analysis code that captures research processes and methodology: the steps taken for tasks like data generation, preparation, analysis, and visualization.
	\item \emph{Novel Methods and Models} captured as software. The results are prototype tools that demonstrate a new idea, method, or model for research.
	\item \emph{Accepted Methods and Models} captured as software. The result can become research software infrastructure that captures more broadly accepted and used ideas, methods, and models for research.
\end{enumerate}
\autoref{fig:SoftwareCategorization}, right, shows the resulting maturity-based categorization.

This categorization system is not yet complete. To illustrate the gap between existing research categories, we characterize them in the following section, based on the characteristics from Table~\ref{table:categorization-criteria}.
Adding more dimensions and refining the dimensions is subject to future work.

\section{Related Research Software Categories}

Research software has been categorized in different contexts to serve different aims.
Some of them are discussed here as related works, as they 
a) represent a good starting point for a discussion on research software categorization, and
b) may be used to compare and assess our categorization.
We characterize these categories via the previously introduced template in the appendix (supplement).

\subsection{Role-Based Categorization}

Van Nieuwpoort and Katz~\cite{NieuwpoortKatz2023} present a role-based categorization.
They categorize research software as an integral component of instruments used in research, as the instrument itself, for analyzing research data, for presenting research results, for assembling or integrating existing components, as infrastructure or an underlying tool, and for facilitating research-oriented collaboration. 
This categorization inspired our work, but they suggest a different set of categories.

\subsection{Maturity-Based Categorization}

In their National Agenda for Research Software~\cite{ARDC2022}, the Australian Research Data Commons -- an Australian research data infrastructure facility -- argue for research software to be recognized as a first-class output of research. They describe a three-level categorization of research software that we adopted for our maturity dimension.

Each category faces specific challenges with regard to recognition, from making research practice transparent, to creating impact through quality software and safeguarding longer-term maintenance. 

\subsection{Application classes in institutional software engineering guidelines}

Institutional guidelines typically define so-called application classes for research software, which require appropriate quality properties, and, thus software engineering methods~\cite{DLR2018}\weg{\cite{FZJ2022}}:
\begin{itemize}
    \ieeeitem For software in Application Class 0, the focus is on personal use in conjunction with a small scope.
    \ieeeitem For software in Application Class 1, it should be possible, for those not involved in the development, to use
it to the extent specified and to continue its development.
	  \ieeeitem For software in Application Class 2, it is intended to ensure long-term development and maintainability. It is
the basis for a transition to product status.
	\ieeeitem For software in Application Class 3, it is essential to avoid errors and to reduce risks. This applies in particular
to critical software.
\end{itemize}
The application classes relate to our maturity domain and to some extent to our developer-based categorization.

\subsection{EOSC Research Software Lifecycle}

The European Open Science Cloud (EOSC) aims to create a virtual environment for sharing and accessing research data across borders and scientific disciplines. 
The SubGroup 1 ``On the Software Lifecycle'' of the EOSC Task Force ``Infrastructure for quality research software'' provides a categorization for software in the research lifecycle~\cite{Courbebaisse_2023}:
\begin{enumerate}
	\item Individual creating research software for own use (e.g. a PhD student).
	\item A research team creating an application or workflow for use within the team.
	\item A team / community developing (possibly broadly applicable) open source research software.
	\item A team or community creating a research service.
\end{enumerate}
This categorization is covered by our developer-based categorization.

\subsection{Computational research in the earth system sciences}

D\"oll et al.~\cite{FIDGEO2023} provide recommendations for sustainable research software for high-quality computational research in the Earth System Sciences, and categorize this research software as follows:
\begin{itemize}
    \ieeeitem Simulation of Earth system processes by Earth system models.
    \ieeeitem Design, processing and analysis of Earth observation and lab experiment data.
	  \ieeeitem Integrative analysis of simulation models, large data bases, and stakeholder knowledge.
\end{itemize}
These categories correspond to our role-based categories~1.1), 1.2), and 1.3), respectively.

\subsection{Categorizing the Software Stack}

Another dimension is the research software stack, from non-scientific infrastructure, scientific infrastructure, discipline-specific software, up to project-specific software~\cite{Hinsen2019}.
This dimension could be the basis for another branch in our multi-dimensiomal categorization.

\section{CONCLUSION}

We categorize research software along various dimensions, contributing to fostering effective development, recognition, and utilization of research software within the research community. One essential use case of this categorization is its incorporation into forthcoming guidelines for research software development. As we classify research software, we enable tailoring guidelines to specific classes, offering developers a structured framework that aligns with each category's unique requirements and challenges.

Moreover, the categorization is intended to be a valuable tool for stakeholders, especially research software engineers and their group, chair, department, or institute leaders. The categorization may provide these individuals with a better understanding of the software they are developing, offering insights into its nature, purpose, and potential impact. This knowledge is essential for informed decision-making, adequate resource allocation, and strategic planning within research institutions.

Recognition for research software engineers is another outcome we anticipate from categorizing research software. By delineating different types of software and acknowledging the diverse skill sets required for their development and maintenance, our categorization aims to contribute to elevating the status of research software engineers. We hope this recognition motivates individuals and fosters a culture that values and appreciates the crucial role played by software in advancing research efforts.

Categorizations may also help assess external software when considering its use. We envision that it contributes to a standardized framework for evaluating software's relevance, applicability, and quality, facilitating informed decisions in adopting tools from different sources.

The categorization may become particularly valuable in allocating project-based or permanent funding. It can help researchers and developers clearly articulate their software's significance in a funding proposal. We envision this classification providing a framework that helps researchers and funding agencies.

Additionally, the categorization may help to emphasize which software is critical, highlighting the importance of its maintenance and continued development for its continued functionality. By highlighting this importance, we seek to contribute to an enhanced awareness of the ongoing support and resources required to ensure the longevity and sustainability of research software.

In the realm of Research Software Engineering (RSE) research~\cite{RSER2023}, we hope that the categorization provides a framework for classifying research objects, supporting software corpus analyses, and enhancing our understanding of the different types of research software and their properties. This structured approach may aid in organizing and interpreting the vast landscape of research software, contributing to advancements in RSE methodologies and practices.

We propose a multi-dimensional categorization of research software, along the dimensions of roles, developers, and maturity. The various dimensions of the categorization are not completely independent of each other. Looking at the dependency between the dimension and identifying constraints on combinations of the dimensions is the subject of future work. Additional dimensions could be the reuse scenarios (such as single-use/single-purpose, extensibility, reusability), the users (such as scientists, humans as research subjects, and citizens), the research software stack~\cite{Hinsen2019}, and the criticality (for instance, mission-critical software). Such extensions and refinements are subject to future work.

\newpage
\onecolumn
\appendix

\textbf{\large Characterization of Related Research Software Categories}

\bigskip

The related research software categories are characterized in terms of the template in \autoref{table:categorization-criteria}.

\autoref{table:Role-based} characterizes the role-based categorization by van Nieuwpoort and Katz~\cite{NieuwpoortKatz2023}.

\autoref{table:ARDC-categorization} characterizes the ARDC categorization.

\autoref{table:guidelineapplicationclasses} characterizes the institutional guideline application class categorization.

\autoref{table:EOSC} characterizes the EOSC research software lifecycle categorization.

\autoref{table:FIDGEO} characterizes the categories in computational research in the Earth system sciences.

\autoref{table:softwarestack} characterizes the software stack categorization~\cite{Hinsen2019}.

\bigskip

\begin{table*}[hb]
    \centering
    \begin{tabularx}{\textwidth}{l X}
    \toprule
        Criterion & Explanation \\
    \midrule
        Scope & Role-based categorization. \\
        Purpose & Funding organizations joined forces to explore how they could effectively contribute to making research software sustainable.\\
        Context & International workshop in 2022 on the future of research software, organized by the Research Software Alliance (ReSA) and the Netherlands eScience Center. \\
        Properties & The roles for research software are defined from the point of view of a researcher, with the goal of making this understandable for funders and policymakers.\\
        Consequences for Creation & Depending on its role category, software is expected to meet different quality requirements and
follow different development processes.\\
      Consequences for Use & Perceive that there are many different types of research software, fulfilling many different roles and functions.\\
        Inter-categorial relations & Individual research software may change its role or take multiple roles.\\
    \bottomrule
    \end{tabularx}
    \caption{Characteristics of the role-based categorization by van Nieuwpoort and Katz~\cite{NieuwpoortKatz2023}.}
    \label{table:Role-based}
\end{table*}

\begin{table*}[hb]
    \centering
    \begin{tabularx}{\textwidth}{l X}
    \toprule
        Criterion & Explanation \\
    \midrule
        Scope & The categorization in~\cite{ARDC2022} supports a discussion about recognition of software in research, with the aim to increase this recognition. \\
        Purpose & The categorization aims to describe the purpose of the software it categorizes as capturing applied or widely accepted research ideas, methodology, and models, or demonstrating new ones. \\
        Context & The categorization has been produced in the context of ARDC's research software policy. \\
        Properties & The properties of the categories represent different challenges faced by software that fall in the respective category. \\
        Consequences for Creation & Depending on its category, software is expected to meet different requirements. 
While analysis code should be FAIR~\cite{FAIR4RS2022}, prototype tools should exhibit a ``high quality'', and research software infrastructure must be created for sustainability, which is realized through safeguarding its long-term maintenance. \\
        Consequences for Use & Software use is featured only implicitly in the categorization.
We expect that software under the different categories are expected to be used differently:
Analysis tools are used for specific research tasks, and are more likely to have a small scope, e.g.,
are applied only to answer a specific research question.
Prototype tools are used to test the methodological hypotheses they implement,
but may also be used experimentally to answer specific research questions.\\
        Inter-categorial relations & The categories are related through \textit{evolution} and \textit{transitive value}. One category evolves from another, e.g., analysis code may evolve into a prototype tool,
that in turn evolves into research software infrastructure.\\
    \bottomrule
    \end{tabularx}
    \caption{Characteristics of ARDC's research software categorization.}
    \label{table:ARDC-categorization}
\end{table*}

\begin{table*}[bt]
    \centering
    \begin{tabularx}{\textwidth}{l X}
    \toprule
        Criterion & Explanation \\
    \midrule
        Scope & Guidelines for software engineering at an academic institution. \\
        Purpose & Identify suitable quality requirements.\\
        Context & Institutional policy and practice. \\
        Properties & Criticality, institutional risk, projected use, development timeline, distribution, commercial exploitation. \\
        Consequences for Creation & Increasingly employ established software engineering methods.\\
        Consequences for Use & Increased (critical) use by increasingly large community. \\
        Inter-categorial relations & Transitive requirements, legal requirements.\\
    \bottomrule
    \end{tabularx}
    \caption{Characteristics of institutional guideline application classes.}
    \label{table:guidelineapplicationclasses}
\end{table*}

\begin{table*}[bt]
    \centering
    \begin{tabularx}{\textwidth}{l X}
    \toprule
        Criterion & Explanation \\
    \midrule
        Scope & Developer- and stakeholder-based categorization. \\
        Purpose & Achieve a common understanding of the current processes in research software engineering, particularly the research software lifecycle.\\
        Context & SubGroup 1 ``On the Software Lifecycle'' of the EOSC Task Force ``Infrastructure for quality research software''. \\
        Properties & Different levels of adopting software engineering practice, different publication requirements and usage scenarios, different stakeholders\\
        Consequences for Creation & Depending on its developer category, software is expected to meet different quality requirements and
follow different development processes.\\
     Consequences for Use & Increasing maturity and support for reproducibility \\
        Inter-categorial relations & Not specified \\
    \bottomrule
    \end{tabularx}
    \caption{Characteristics of the EOSC research software lifecycle categorization.}
    \label{table:EOSC}
\end{table*}

\begin{table*}[bt]
    \centering
    \begin{tabularx}{\textwidth}{l X}
    \toprule
        Criterion & Explanation \\
    \midrule
        Scope & Recommendations for universities, funders, and the scientific community. \\
        Purpose & Safeguard the quality and efficiency of computational research in Earth System Sciences and make research results that have been generated by research software reproducible.\\
        Context & Ideas of a DFG round table meeting on sustainable research software for high-quality computational research in the Earth System Sciences. \\
        Properties & Research software developed in the Earth System Sciences is characterized by the complexity of the underlying models, multifaceted dependencies, the multi-modality of the data, and the size of the data, which can impose specific hardware and software requirements.\\
         Consequences for Creation & Depending on its role category, software is expected to meet different quality requirements and
follow different development processes.\\
    Consequences for Use & Dependency on the research cycle \\
        Inter-categorial relations & Combination, integration \\
    \bottomrule
    \end{tabularx}
    \caption{Characteristics of categories in computational research in the Earth system sciences.}
    \label{table:FIDGEO}
\end{table*}

\begin{table*}[bt]
    \centering
    \begin{tabularx}{\textwidth}{l X}
    \toprule
        Criterion & Explanation \\
    \midrule
        Scope & Describing principles of software collapse. \\
        Purpose & Identify dependent layers of different (academic) specificity to model threat.\\
        Context & Research software sustainability. \\
        Properties & Domain specificity. \\
        Consequences for Creation & Build on stable lower layers, quickly react to threats, accept agility.\\
        Consequences for Use & Decreasing specificity of application domain from top to bottom.\\
        Inter-categorial relations & Dependency, transitive threats.\\
    \bottomrule
    \end{tabularx}
    \caption{Characteristics of categorizing the software stack.}
    \label{table:softwarestack}
\end{table*}

\clearpage

\section*{REFERENCES}

\end{document}